\documentclass[ ]{aa}  

\usepackage{graphicx}
\usepackage{txfonts}
\usepackage[utf8]{inputenc}
\usepackage{comment}
\usepackage{color}
\usepackage{xcolor}
\usepackage{url}
\usepackage{hyperref}
\usepackage{multirow}
\usepackage{float}
\usepackage{orcidlink}
\usepackage{soul} 
\usepackage{ulem} 
\usepackage{bm}
\definecolor{myblue}{HTML}{6a359c} 
\definecolor{myviolet}{HTML}{756BB1}
\definecolor{mygreen}{HTML}{7FCDBB}
\definecolor{myyellow}{HTML}{FFFDD1}
\usepackage{mathtools}
\hypersetup{
	colorlinks,
	linkcolor={myblue},
	citecolor={myblue},
	urlcolor={myblue}}
	
\begin{document} 
\title{Planetary atmospheric escape and disk formation around WDJ0914+1914}
 \titlerunning{Photoevaporation and disk formation}
\authorrunning{Villarreal D'Angelo et al.}

\author{
C. Villarreal D'Angelo\orcidlink{0000-0003-1701-7143}\inst{1},
M. P. Ronco\orcidlink{0000-0003-1385-0373}\inst{2}, 
M. R. Schreiber\orcidlink{0000-0003-3903-8009}\inst{3},
O. Toloza\orcidlink{0000-0002-2398-719X}\inst{3},
A. Esquivel\orcidlink{0000-0001-7222-1492}\inst{4},
B.~T. G\"ansicke\orcidlink{0000-0002-2761-3005}\inst{5}}

\offprints{carolina.villarreal@unc.edu.ar}
\institute{
{Instituto de Astronom\'{\i}a Te\'orica y Experimental (CONICET-UNC). Laprida 854, X5000BGR C\'ordoba, Argentina}
\and
{Instituto de Astrof\'{\i}sica de La Plata, CCT La Plata-CONICET-UNLP, Paseo del Bosque S/N (1900), La Plata, Argentina},
\and
{Departamento de F\'{\i}sica, Universidad T\'ecnica Federico Santa Mar\'{\i}a, Av. Espa\~na 1680, Valpara\'{\i}so, Chile}
\and
{Instituto de Ciencias Nucleares, Universidad Nacional Autónoma de México, Apartado Postal 70-543, 04510,  Ciudad de México, México}
\and
{Department of Physics, University of Warwick, Coventry, CV4 7AL, UK}}

\date{}

 \abstract
   {The spectrum of the white dwarf WD\,J091405.30+191412.25 displays the absorption and double-peaked emission lines of the volatiles hydrogen, oxygen, and sulfur. This unique characteristic has been interpreted as evidence of this white dwarf accreting mass from a circumstellar disk that had formed from atmospheric material evaporating off a close-in Neptune-like or {\it super-puff} mass planet. Thus far, however, the orbital separation of the planet and its mass-loss rate have only been estimated using simple analytical approximations.}
   {In this work, we tested the outlined scenario by determining the mass-loss rate of the irradiated planet through hydrodynamical simulations and by studying in detail the formation and evolution of the disk, as well as the resulting mass accretion rates onto the white dwarf.}
   {We used 3D radiative-hydrodynamic simulations of a pure hydrogen atmosphere under the influence of extreme UV (XUV) radiation from the white dwarf to determine the mass loss rates for Neptune-like and {\it super-puff} planets at different orbital separations. The escaping material was assumed to feed a circumstellar disk, whose viscous evolution was then modeled using 1D radial disk calculations to determine the resulting accretion rates in each case.
   }
   {For the explored cases, the planetary mass loss rates span a narrow range of $\sim (1.8 -4.0)\times10^{12}$ g s$^{-1}$.
   This material forms a disk around the star that stabilizes in $\lesssim 10^5$ years, as a result of the balance between continuous mass injection from the evaporating planet and viscous mass loss onto the central star. The accretion rates at this stage are on the order of a few $\times 10^{9}$–$10^{10}$ g s$^{-1}$, in agreement with previous observational estimates. Our simulations suggest that the disk extends beyond the planet position, in contrast to the size derived from observations.}
   {We conclude that a planet with a gaseous envelope at a distance of $\sim15$ R$_\odot$ will indeed suffer from photoevaporation and that the material escaping the planet will form a gaseous disk around WD\,J0914+1914, most likely extending beyond the location of the planet. Our results confirm the possibility that the observed spectral features of WD\,J0914+1914 can be explained by an evaporating gas-rich planet.}


   \maketitle
%

\section{Introduction}
\nolinenumbers
Clear evidence to support the existence of planetary material around white dwarfs comes from the detection of heavy metals in the atmospheres of a significant fraction of white dwarfs \citep{Koester2014}, as well as the presence of gaseous and dusty debris disks around white dwarfs 
\citep[e.g.][]{Zuckerman2003,Gaensicke2006,Farihi2009}$-$with one potentially hosting an orbiting planetesimal \citep{Manser2019}$-$coupled with the direct observation of transiting disintegrating planetesimals \citep[e.g.][]{Vandenburg2015}. Although direct detections of planets around white dwarfs are still observationally challenging, several exoplanets orbiting white dwarfs have been identified  \citep{Sigurdsson2003,Luhman2011,Gaensicke2019, Vandenburg2020,Blackman2021,Mullally2024,Limbach2024}.  
In some of these cases, the potentially detected planets appear to be very close to the white dwarf \citep{Gaensicke2019,Vandenburg2020,Limbach2024}, which implies that dynamical interactions have played a role in sculpting the current configuration \citep{Vandenburg2020,Munoz2021,Lagos2021}. 

One of the most intriguing cases of a close-in planet around white dwarfs is the potentially evaporating planet orbiting the white dwarf WD\,J091405.30+191412.25 (hereafter WD\,J0914). The existence of this planet was inferred from the detection of oxygen and sulfur absorption lines that indicate the accretion of volatile elements at a rate of $\sim3.3\times 10^9$ g s$^{-1}$ as well as double-peaked hydrogen, oxygen, and sulfur emission lines that provide clear evidence for the existence of an accretion disk surrounding the white dwarf \citep{Gaensicke2019}. The elements and their abundances detected in WD\,J0914 differ from all other cases of white dwarfs accreting planetary material, leading to the interpretation that the hot white dwarf is irradiating a giant planet, resulting in atmospheric mass loss \citep{Gaensicke2019}. As WD\,J0914 is a moderately hot star (T$_\text{eff}= 27\,743$ K), most of its luminosity is concentrated in the extreme ultraviolet (XUV) range ([5-912]\,\AA), resulting in the photoionization of hydrogen and, hence, the evaporation of the planetary atmosphere.  

Following the discovery and interpretation of WD\,J0914b, the hydrodynamic escape of irradiated planets around white dwarfs has been further investigated. \citet{Schreiber2019} showed that accretion from evaporating planets is capable of explaining the photospheric absorption lines of volatile elements, found in one-third of hot single white dwarfs. More recently, \citet{Gallo2024} presented a numerical exploration of the evaporation suffered by a planetary hydrogen-helium atmosphere due to XUV radiation of a hot white dwarf with M=0.6\,M$_\odot$. They used a 1D hydrodynamic code, tested two different gaseous planets at the same orbital distance ($0.02$\,au), and found that the planetary mass-loss rate reaches $10^{14}$ g s$^{-1}$ for white dwarf temperatures above $\sim5\times 10^4$ K and is about $10^{12}$ g s$^{-1}$ for temperatures below $\sim2\times10^{3}$ K. 
The mass-loss rate of the assumed irradiated planet around WD\,J0914 of $\sim5\times10^{11}$ g s$^{-1}$ was estimated using a simple scaling law derived from hydrodynamical simulations of planets around main sequence stars \citep{Gaensicke2019}. Although the value obtained seems to be in the range of those found by \citet{Gallo2024}, we still lack dedicated 3D simulations of the hydrodynamic escape that has been presumed to occur in the atmosphere of the hypothetical planet around WD\,J0914. 

Furthermore, the orbital separation of the potential planet is uncertain. \citet{Gaensicke2019} argued that WD\,J0914b could be located at 15\,R$_\odot$. To reach such a close orbit, the planet must have migrated inward from more distant regions on an extremely short timescale (given the high temperature and the corresponding short cooling age of the white dwarf). The eccentric Kozai-Lidov (EKL) mechanism could explain the current position of WD\,J0914b \citep{Stephan2017, Stephan2021}, but this scenario requires a stellar companion that has not been identified thus far. The original scenario proposed by \citet{Gaensicke2019} was complemented by \citet{VerasFuller2020}, who suggested several alternative dynamical scenarios, including high-eccentricity migration with chaotic tides. 

In this work, we used detailed numerical simulations to investigate the viability of the scenarios proposed for the WD\,J0914 system. We approached the problem by performing 3D radiative-hydrodynamic simulations of the atmospheric escape to estimate the planetary mass-loss rate for different planetary masses and different positions of the planet. The mass-loss rate was then used to investigate the formation and evolution of the accretion disk around WD\,J0914 and to determine the accretion rate in the white dwarf. We generally find that the scenario described by \citet{Gaensicke2019} remains plausible, as the atmospheric mass loss rates and accretion rates we determined here are similar to those derived from observations. However, in contrast to what was assumed in the work of \citet{Gaensicke2019}, the accretion disk formed from the escaping atmosphere is predicted to extend beyond the position of the evaporating planet. 

\section{Observational constraints and a possible dynamical history of the system}\label{obs_constrains}

Before simulating the hydrodynamic escape of the atmosphere of a potential planet around WD\,J0914, we reviewed the currently available observational constraints and the proposed dynamical scenarios.  
WD\,J0914 is a hot, very young, metal-polluted white dwarf with an estimated mass of 0.56\,M$_\odot$, a surface gravity of log\,(g) = 7.85 dex, a radius of 0.015\,R$_\odot$, an effective temperature of 27\,743~K, and a cooling age of approximately 13.3\,Myr. Its spectra exhibit metal absorption lines indicative of ongoing accretion of planetary material. The absence of Zeeman splitting in these lines suggests that the white dwarf is not strongly magnetic  \citep{Wilson2021}.
 
The elements detected in the disk (hydrogen, oxygen, and sulfur) can come from the evaporation of a gaseous planet, similar to the Neptune-like planet HAT-P-26b \citep{Hartman2011}. From the disk extension and by assuming a circular orbit, \citet{Gaensicke2019} assumed the orbital separation of the planet to be $\sim$15\, R$_\odot$ (0.07 au). This estimated position of the planet is hard to explain from a dynamical point of view, as the cooling age of the white dwarf is comparable to the circularization timescale of a scattered planet. This was investigated by \citet{VerasFuller2020} who studied different dynamical scenarios that could lead to the estimated planet position reported in \citet{Gaensicke2019}. 
Two of the explored scenarios provide an initial condition for our atmospheric escape simulations. The first (and likely the more viable) scenario is based on the assumption that the planet is a so-called super-puff planet, which would allow for chaotic tidal forces acting to shrink and circularize the orbit of the planet with an orbital distance of $a_{\text{p}}$=0.07\,au. The second scenario assumes that the mass of the planet is $<20$ M\,$_\oplus$ and orbits within the disk or at its outer edge, around $a_{\text{p}}\sim$0.04\,au. At this closer separation, tidal circularization can act as well within the short cooling age of the white dwarf. The other scenarios suggested in this work, such as a planet in a highly eccentric orbit, are not suitable for the numerical model used here due to the high computational costs they entail.

Since the position and the type of planet (i.e., super-puff or Neptune-like) influence the planetary atmospheric mass loss, we explored three different scenarios, namely: the two mentioned above taken from \citet{VerasFuller2020} and the one originally proposed in \citet{Gaensicke2019}. For each of the cases we determine the planetary mass-loss rate and follow the evolution of the resulting disk for several hundred thousand years to estimate the accretion rate onto the white dwarf.

\section{Methods and initial conditions}\label{method}

To investigate the escaping atmosphere of a gaseous planet due to the intense XUV radiation of WD\,J0914 along with the long-term evolution of the potential disk formed from this material, we employed two different numerical approaches. First, we used 3D radiative-hydrodynamic simulations with the publicly available code {\sc guacho}\footnote{https://github.com/esquivas/guacho} to compute the planetary mass-loss rate. Then, using the mass-loss rate derived from the 3D models, we employed 1D viscous disk evolution simulations using the PLANETALP code \citep{Ronco2017,Guilera2020} to model the behavior of the escaped material over periods of hundreds of years and to compute the stellar accretion rates.

Our adoption of a two-step approach to study the WD\,J0914 system is motivated by the high computational cost of fully self-consistent global 3D simulations that include the physics of atmospheric escape alongside the formation and long-term (hundreds of years) evolution of the circumstellar disk. Therefore, we used 3D models to compute atmospheric escape from first principles and to capture the spatial distribution of the escaping material, allowing for a more robust estimate of the planetary mass-loss rate. We then used faster 1D models to follow the long-term evolution of the escaping material over hundreds of years and to compute its accretion rate onto the central star.

We modeled three different scenarios for the WD\,J0914 system. The first one adopts the orbital separation and planet characteristic presented in \citet{Gaensicke2019}, a Neptune-like planet with the same mass and radius as HAT-P-26b at an orbital distance of 0.07\,au, named {\it a$_\text{p}$007}. The other two were taken from the work of \citet{VerasFuller2020}. Model {\it a$_\text{p}$004} describes a planet that resembles HAT-P-26b, with a mass that is lower than 20 M$_\oplus$, at an orbital distance of 0.04\,au, while model {\it super-puff} describes a planet with a mass and radius resembling that of a super-puff planet, at a distance of 0.07\,au from the star. In all cases, the planet is assumed to be in a circular orbit around the star. 

\begin{figure}
   \centering
   \includegraphics[width=0.9\linewidth]{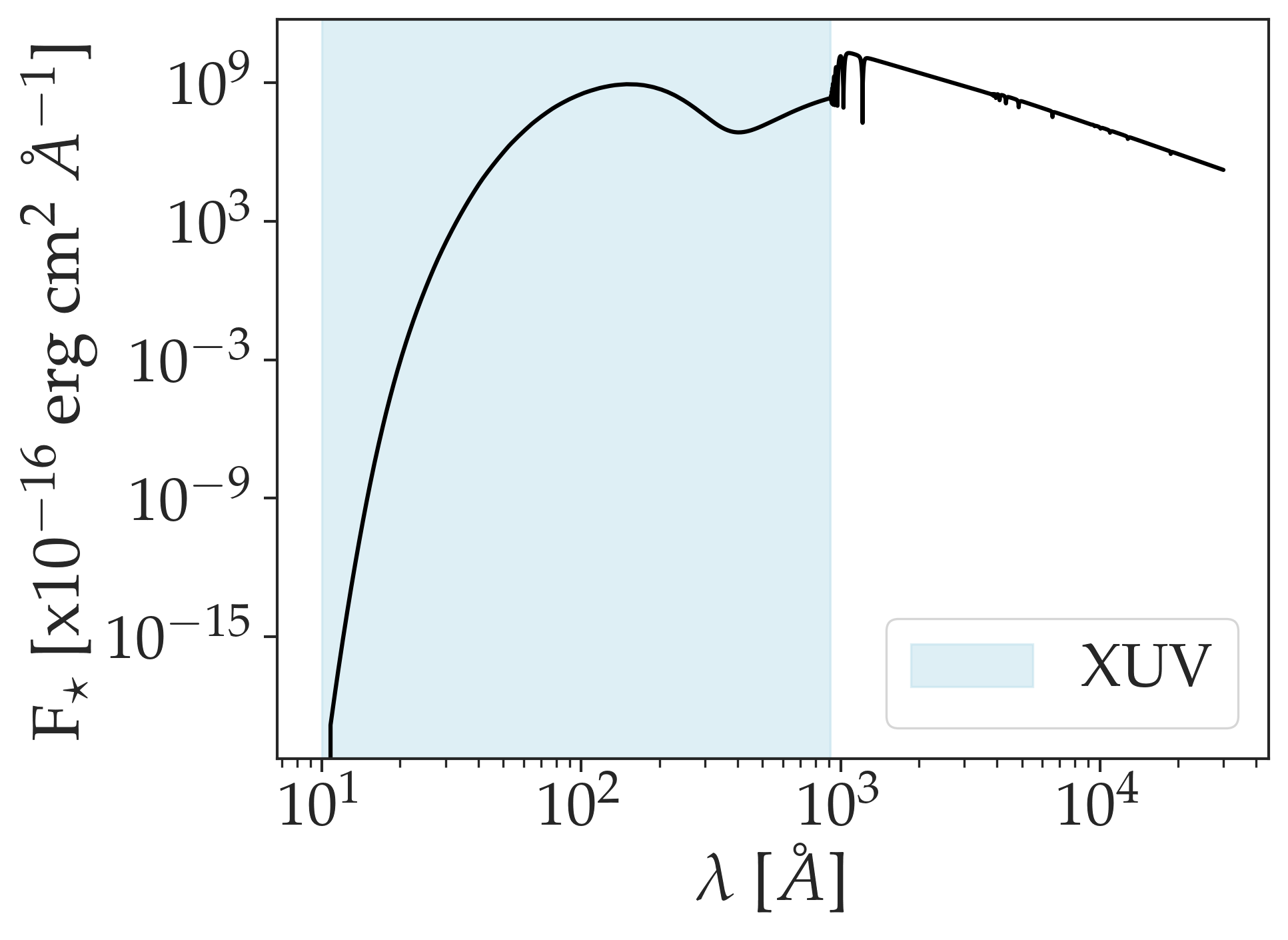}
    \caption{Stellar flux of WD\,J0914 from the atmospheric model of \citet{Koester2010} used to derive the XUV flux for 3D models.} 
   \label{fig:stellar_spec}
\end{figure}

The stellar parameters were taken from \cite{Gaensicke2019}, while the stellar flux in the XUV range (F$_\text{XUV}$), computed at the planet position, is derived using the model atmosphere code of \cite{Koester2010} and is presented in figure \ref{fig:stellar_spec}. The physical and orbital characteristics of the planet for each model are presented in Table \ref{tab:1}.
\begin{table}
\caption{Physical and orbital properties of the planets considered for our simulations, following \citet{Gaensicke2019} and \citet{VerasFuller2020}.}
    \centering
    \renewcommand{\arraystretch}{1.5} 
    \begin{tabular}{|l|c|c|c|}
     \hline 
     & {\it a$_{\text{p}}$004} & {\it a$_{\text{p}}$007} & {\it super-puff} \\
      \hline
      M$_{\text{p}}$[$M_\oplus$] &  18 & 18 & 6 \\
      \hline
      R$_{\text{p}}$[R$_\oplus$] & 7 & 7 & 4 \\
      \hline
      F$_{\text{XUV}}$ [$\frac{\text{erg}}{\text{cm}^2 \text{s}}$] & 1.3$\times 10^6$  & 4.3$\times 10^5$ & 4.3$\times 10^5$ \\
      \hline
      $a_{\text{p}}$~[au] &  0.04 & 0.07 & 0.07 \\ 
      \hline
      T$_\text{orb}$ [day] & 3.9 & 8.9 & 8.9\\
        \hline
    \end{tabular}
    \vspace{0.2cm}
    \tablefoot{ 
    The mass and radius considered for models {\it a$_\text{p}$004} and {\it a$_\text{p}$007} correspond to those of HAT-P-26b, while for model {\it super-puff}, they have been adopted from the values in \citet{VerasFuller2020}. The stellar flux is computed at the planet's position.} 
    \label{tab:1}
\end{table}   

\subsection{3D atmospheric escape simulations}
\label{apA}

The 3D code {\sc guacho} that we employed for our simulations has already been used in numerous works to study different phenomena, such as the star-planet wind-wind interaction \citep{Schneiter2016,Villarreal2018,Esquivel2019,Villarreal2021}, the photoevaporation of planetary atmospheres due to stellar radiation \citep{Sgro2022}, and the accretion of planetary material around a white dwarf \citep{Estrada-Dorado2024}. The code solves the set of the hydrodynamical equations of mass, momentum, and energy conservation for a single fluid in a uniform cartesian coordinate system, with a second-order accurate Godunov-type method, using a linear slope-limited reconstruction and the Harten-Lax-van Leer-Contact (HLLC) approximate Riemann solver \citep{Toro1994}.

We assume that the radiative energy gains are only due to the photoionization of neutral hydrogen and, for a monochromatic flux, these values can be obtained with $n_\mathrm{HI} \phi \epsilon$, where $n_\mathrm{HI}$ is the number density of neutral hydrogen, $\phi$ is the photoionization rate, and $\epsilon= (h\nu - h\nu_0)$ is the average energy gain per photoionization, where $\nu_0=13.6/h$ eV is the hydrogen threshold frequency and $h$ is the Planck constant. 

Radiation losses can be calculated using the analytical prescription presented in \cite{Biro1995}, used in \citet{Villarreal2021} and \citet{Sgro2022}. The nonequilibrium cooling function, valid for temperatures above $1\times10^4$ K, includes collisional ionization, collisional excitation of the Lyman-$\alpha$ line, and recombination. We also included cooling due to the collisional excitation of the forbidden O$_\mathrm{I}$ and O$_\mathrm{II}$ lines, assuming that the ionization fraction of O$_\mathrm{II}$ follows that of H$_\mathrm{II}$ and is multiplied by a factor of $A=7.033$, to account for the contribution of other important ions that produce cooling such as C, N, and S (see \citealt{Biro1995} for details). Above T=$5\times10^4$ K, the cooling function adopts the cooling of an ionized gas in coronal equilibrium. 

The composition of the planetary atmosphere was assumed to be pure neutral hydrogen, the most abundant species in this type of planet. Heavier elements found in the observations by \citet{Gaensicke2019} are not included explicitly in the model, but their effect, attributed to radiative cooling, was taken into account, as mentioned above.

Together with the hydrodynamic equations, we solved a rate equation for neutral hydrogen via
\begin{equation}
 \begin{split}
  \frac{\partial n_\mathrm{HI}}{\partial t} + \nabla.(n_\mathrm{HI}
  {\bf{u}})=  & (n_\mathrm{H}-n_\mathrm{HI})^2\alpha(T)\\
& -(n_\mathrm{H}-n_\mathrm{HI})n_\mathrm{HI}c(T)-n_\mathrm{HI}\phi,
 \end{split}
\label{eq:hrate}
\end{equation}
where $n_\mathrm{H}=\rho/m_\mathrm{H}$ is the total hydrogen number density (with $m_\mathrm{H}=1.66\times10^{-24}\,\mathrm{g}$ the proton mass) and $n_\mathrm{HI}$ is the number density of neutral hydrogen. Then, $\alpha(T)=2.55\times10^{-13}(10^4/T)^{0.79}$, $c(T)=5.83\times10^{-11}\sqrt{T}\exp(-157828/T) $ are the "case B" recombination and collisional ionization rates taken from \cite{Osterbrock1989}. 

The photoionization rate was calculated within each cell using the formula $\phi = \mathrm{S}/(n_\mathrm{HI}dV)$, where $\mathrm{S}=\mathrm{S}_0\exp(-n_\mathrm{HI}a_0\mathrm{dl})$ is the ionizing photon-rate that is attenuated by the absorbing material within the cell of volume dV. Then, $a_0=6.3\times10^{-18}$ cm$^{2}$ is the photoionization cross-section of hydrogen at the threshold frequency and dl is the path that the photon travels, taken in steps of one half of the cell size. Finally, $\mathrm{S}_0$ is the initial photon number at the planet position in the XUV range estimated from integrating the model spectrum of WD\,J0914 and divided by the photon energy. Assuming a monochromatic flux, the representative energy of each photon in the XUV is taken to be 15 eV. Assuming that all the energy in photoelectrons goes to heat the gas, the total energy of stellar photons is the energy absorbed in a photoionization (13.6 eV) plus the energy gained by the photoelectrons which is calculated as the ratio of the heating rate over the photoionization rate. 

In the simulation, the stellar ionizing radiation was grouped into $10^{7}$ photon packages that were launched plane-parallel from one of the mesh faces. These were traced within the mesh using the ray-tracing method described in \citet{Esquivel2013}, later adapted for the photoionization of exoplanet atmospheres in \citet{Schneiter2016}. This approach allowed us to follow the ionization state of hydrogen (and thus O$_\mathrm{II}$)\footnote{This approximation is justified by the efficient charge exchange between hydrogen and oxygen for temperatures below $5\times10^4$ K. At higher temperatures O$_\mathrm{III}$ becomes more abundant, and thus we switch to a coronal equilibrium cooling curve.}, which serves as an input for the radiative cooling calculation.

\subsection{Radiation pressure by Lyman-$\alpha$}
\begin{figure}
    \centering
    \includegraphics[width=\linewidth]{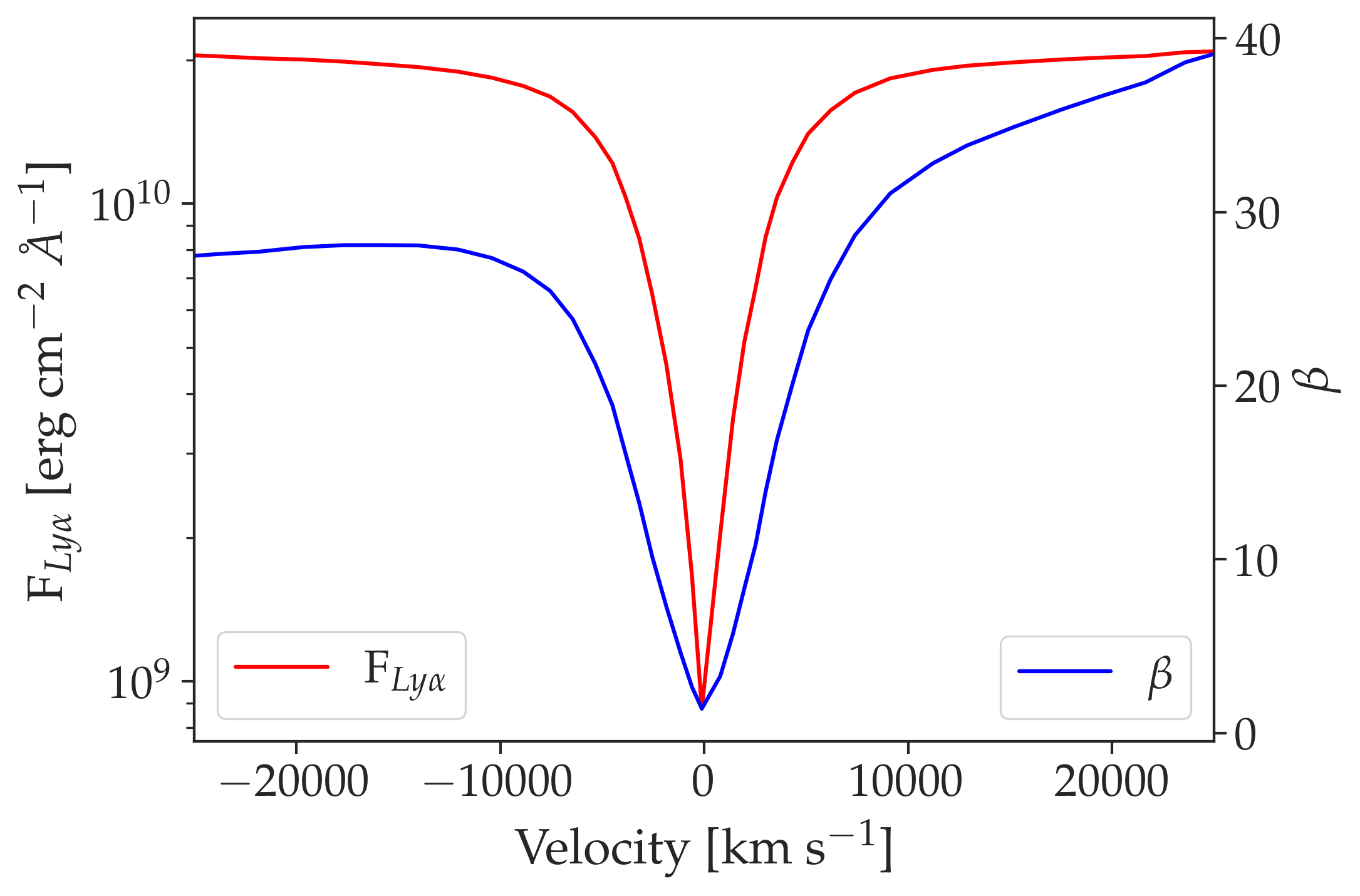}
    \caption{Stellar flux in Lyman-$\alpha$ from the stellar atmospheric model of \cite{Koester2010} and the $\beta$ parameter calculated using equation (2) of \citet{Lagrange1998}. }
    \label{fig:beta_parameter}
\end{figure}

To include the effect of radiation pressure from Lyman-$\alpha$ photons in our 3D models, we followed the approach of \cite{Esquivel2019}. We computed the ratio of radiation pressure to stellar gravity, $\beta$, as defined in \cite{Lagrange1998}. The effective stellar gravity acting on neutral hydrogen was then $(1 - \beta(v)\chi_n)g_\star$, where $\chi_n$ is the neutral fraction, and $g_\star$ the stellar gravitational acceleration. The parameter $\beta$ depends on the stellar flux in the line profile and varies with the line-of-sight velocity of the absorbing material.
The resulting $\beta$ values, derived using the Lyman-$\alpha$ flux from the stellar spectrum model of \cite{Koester2010}, are shown in figure \ref{fig:beta_parameter}. For WD\,J0914, the strong Lyman-$\alpha$ flux yields $\beta > 1$, indicating that radiation pressure exceeds stellar gravity, indicating the importance of this effect, as noted in \citet{Gaensicke2019}.

\subsection{3D simulation setup}
To study the planetary atmospheric escape for each of our models, we used the numerical setup employed in \cite{Sgro2022, McCann2019}. We positioned the planet at the center of the grid and solved the hydrodynamic equations in time in a noninertial frame that moves together with the orbital motion of the planet. The frame was centered at the barycenter of the star-planet system. For simplicity, we neglected the planetary rotation, namely, we assumed it to be tidally locked.

The physical extension of the mesh was 40 planetary radii, discretized with $1000\times256\times1000$ total cells in the $x$, $y$, and $z$ directions, respectively. This provides a resolution of 0.04\,R$_\text{p}$. The coordinate system was chosen so that the orbital plane would remain in the $x-z$-plane and the planet would move in the $-x$ direction. The star, although it is not within the mesh, was located in the $-z$ direction.

Our initial setup consisted of a planetary core of constant density for r < 0.5 R$_\text{p}$, surrounded by three concentric isentropic atmospheres that are pressure-matched at their interfaces. The first two isentropic layers represent the planetary atmosphere (0.5 R$_\text{p}$< r < R$_\text{e}$), where R$_\text{e}$ > 1 R$_\text{p}$, is the outer boundary of the atmosphere, set near the radius at which the hydrostatic density profile approaches zero. Within this region, density and pressure were determined from the hydrostatic equilibrium solution, using the initial temperature and density specified at 1 R$_\text{p}$. The third outermost layer corresponds to the low-density ambient medium. Its pressure was matched to that of the planetary atmosphere at R$_\text{e}$, preventing the collapse of the ambient gas onto the planet. Boundary conditions were fixed for r < 0.75 R$_\text{p}$ at every time step. 
This setup follows the approach described in \cite{McCann2019} and we refer to that work for further details. 
We chose a temperature of $1\times10^3$ K and a density of $1\times10^{-13}$ g cm$^{-3}$. With this density, we were able to ensure that the main absorption region of XUV photons, defined by an optical depth of $\tau = 1$, would lie close to 1 R$_{\text{p}}$ at the beginning of the simulation. This region corresponds to where the atmosphere is most strongly heated and where the outflow is launched. The $\tau=1$ surface eventually changes in this setup, since the planetary outflow ends up finding its base self-consistently as the simulation evolves, but it does lie close to this initial approximation. Changing the initial values of temperature and density will not change the solution of the planetary outflow as long as the temperature at the base remains below $1\times10^4$ K and the density is such that inside the planet, $\tau>1$, as discussed in \citet{Murray-Clay2009}.

For every model, we computed the instantaneous planetary mass-loss rate every 500 simulation steps. This instantaneous mass-loss rate is calculated by considering the difference in the total mass inside a sphere of radius 2\,R$_{\text{p}}$, centered at the position of the planet, before and after advancing the solution a single time-step (and divided by the corresponding $\Delta t$). The adopted mass-loss rate, $\dot{M}_\text{p}$, is the instantaneous value averaged over the last 20 values.

\subsection{1D disk viscous evolution simulations}
\label{sec:1Dmodel}
After computing the planetary mass-loss rate with our 3D models, we studied the behavior of this material over secular timescales (up to $10^5$ years) to estimate mass accretion rates onto the white dwarf. For this purpose, we used the 1D isothermal disk evolution module of the PLANETALP code, adapted to our white dwarf disk-planet scenario.

The code computes the viscous evolution of an axisymmetric gaseous disk characterized by a gas surface density profile, $\Sigma_{\text{g}}$, formed, in this case, from the evaporated planetary atmosphere. The evolution of $\Sigma_{\text{g}}$ is represented by a diffusion equation \citep{LyndenBell1974,Pringle1981} that includes the tidal torque term exerted by the planet \citep{LinPapaloizou1986,Alexander2011,NayakshinLodato2012, Ronco2021} via
\begin{align}
  \frac{\partial \Sigma_{\text{g}}}{\partial t}= & \frac{3}{R}\frac{\partial}{\partial R} \left[ R^{1/2} \frac{\partial}{\partial R} \left( \nu \Sigma_{\text{g}} R^{1/2}  \right) - \frac{2\Sigma_{\text{g}} \Lambda}{3\Omega} \right] + \dot{\Sigma}_{\text{g}}(R), 
\label{eq:evol_gas}
\end{align}
where $R$ is the distance to the central star and $\nu= \alpha c_\text{s} \text{H}_{\text{g}}$ is the kinematic viscosity, with $\alpha$ being the dimensionless viscosity parameter from \citet{Shakura1973}. Then, $c_\text{s}$ the sound speed and $\text{H}_{\text{g}}=c_\text{s}\Omega^{-1}$ is the scale height of the disk with $\Omega$ as the Keplerian frequency. The sound speed is given by $c_\text{s}= \sqrt{\frac{\gamma k_{\text{B}} \text{T$_g$}}{\mu m_\text{H}}}$,
where $\gamma=5/3$ is the ratio between specific heat capacities, $k_{\text{B}}$ is the Boltzmann-constant, T$_g$ is the gas disk temperature, $\mu$ is the mean molecular weight that depends on the gas abundances$-$ in our case, set to 10 following \citet{Gaensicke2019}$-$and $m_\text{H}$ is the mass of a hydrogen atom. Next, $\Lambda$ is the specific tidal torque that represents the amount of angular momentum transferred by the planet to the disk at a radius, $R$, per unit time and disk mass, given by \citep{ArmitageNatarajan2002,NayakshinLodato2012} via
\begin{equation}
\Lambda(R) =
\begin{dcases}
\frac{f}{2} q^2 \Omega^2 R^2 \left(\frac{a_{\text{p}}}{\Delta}\right)^4, & R > a_{\text{p}}, \\
-\frac{f}{2} q^2 \Omega^2 R^2 \left(\frac{R}{\Delta}\right)^4, & R < a_{\text{p}}.
\end{dcases}
\label{eq:Torque}
\end{equation}
where $f$ is a dimensionless normalization parameter between 0.001 and 1 \citep{ArmitageNatarajan2002,Ronco2021}, $q$ represents the mass ratio between the planet and the white dwarf, 
and $\Delta = {\text{max}}(R^{\text{p}}_{\text{Hill}},{\text{H}}_{\text{g}},|R-a_{\text{p}}|)$, with $R^{\text{p}}_{\text{Hill}}=a_{\text{p}}(q/3)^{1/3}$ as the Hill radius of the planet.

Finally, $\dot{\Sigma}_{\text{g}}$ acts as a source term and represents the mass deposition associated with planetary atmospheric escape. Numerically, the gas is injected following a Gaussian radial profile centered at the planetary orbit and normalized such that $\dot{M}_{\text{p}}=\int{2\pi R\dot{\Sigma}_\text{g}dR}$ (i.e., the total injected mass equals the planetary mass-loss rate).

As in \citet{Kenyon&Bromley2017}, we adopted a simple power law for the gas disk temperature given by 
\begin{eqnarray}
  T_g = T_0\left(\frac{R}{R_\odot}\right)^{-n},
\end{eqnarray}  
where $T_0=10000$ K and $n=0.80$ were determined by fitting the results from \citet{Melis2010} for a white dwarf with $T_\text{eff}=20000$ K and scaling them for $T_\text{eff}=30000$ K, a value closer to the effective temperature of WD\,J0914. With these parameters, the resulting disk temperatures range between $\sim10000$ K and $\sim3000$ K, as suggested by \citet{Gaensicke2019}.

To account for the absorption of XUV photons as the disk grows, which can diminish atmospheric escape from the planet, we assumed that the planetary mass-loss rate has a dependence on the optical depth. This takes the form of $\dot{m}_{\text{p}}(t)=\dot{M}_{\text{p}}\exp(-\tau)$, with $\dot{M}_{\text{p}}$ as the mass-loss rate derived from 3D models. Here, $\tau$ is calculated as the ratio of the column density of neutral hydrogen over the density at which the medium becomes optically thick, taken to be $5\times10^{17}$ cm$^{-2}$ \citep{Murray-Clay2009}. The fraction of neutral hydrogen ($x_\text{HI}$) within the disk was assumed to remain constant to a given value taken from the results of the 3D models. The neutral hydrogen density was then derived by multiplying this fraction by the total surface density in the disk. 

The simulation was initialized with a zero gas surface density across the entire radial grid, extending from $0.1\,R_\odot$ to $100\,R_\odot$. As the planet begins to lose mass due to the irradiation from the white dwarf, this material is injected into the model leading to the gradual formation of a gaseous disk. Its viscous evolution is then computed by solving equation \ref{eq:evol_gas} at every time step using 1000 logarithmically spaced radial bins. Once the gas reaches the inner edge of the grid, the mass accretion rate to the white dwarf can be calculated as $\dot{M}_{\text{acc}} = -2\pi R\Sigma_{\text{g}}v_{\text{R}}$, with $v_{\text{R}}$ as the gas radial velocity.

\begin{table}
\caption{Results from our numerical models}
    \centering
    \renewcommand{\arraystretch}{1.5} 
    \begin{tabular}{|c|c|c|c|}
     \hline 
     & {\it a$_{\text{p}}$004} & {\it a$_{\text{p}}$007} & {\it super-puff} \\
      \hline
      ${\dot M_{\text{p}}}$ [g~s$^{-1}$] &  $4.01\times10^{12}$ & $2.04\times10^{12}$ & $1.80\times10^{12}$\\
      \hline
      M$_{\text{d}}$ [g] & $8.4\times10^{21}$&  $4.01\times10^{21}$& $3.93\times10^{21}$  \\ 
      \hline
      ${\dot{M}_{\text{acc}}}$ [g~s$^{-1}$] & $1.36\times10^{10}$ & $4.5\times10^{9}$ & $4.25\times10^{9}$  \\
      \hline
    \end{tabular}
    \vspace{0.2cm}
    \tablefoot{${\dot M_{\text{p}}}$ represents the mass loss rates computed for each case with our 3D hydrodynamical simulations, while M$_{\text{d}}$ and ${\dot M_{\text{acc}}}$ represent the disk mass and the accretion rate onto the white dwarf, respectively, computed with our 1D model, once the disk reaches a steady state.}
    \label{tab:2}
\end{table}

\section{Results and analysis}\label{results}

\subsection{3D atmospheric escape simulations: Mass loss rates }\label{Res3DLocal}
We modeled the response of a neutral hydrogen planetary atmosphere to the incoming stellar XUV flux from WD\,J0914 for the three scenarios presented in Table \ref{tab:1}.
\begin{figure}[ht!]
\begin{center}
\includegraphics[width=0.9\linewidth]{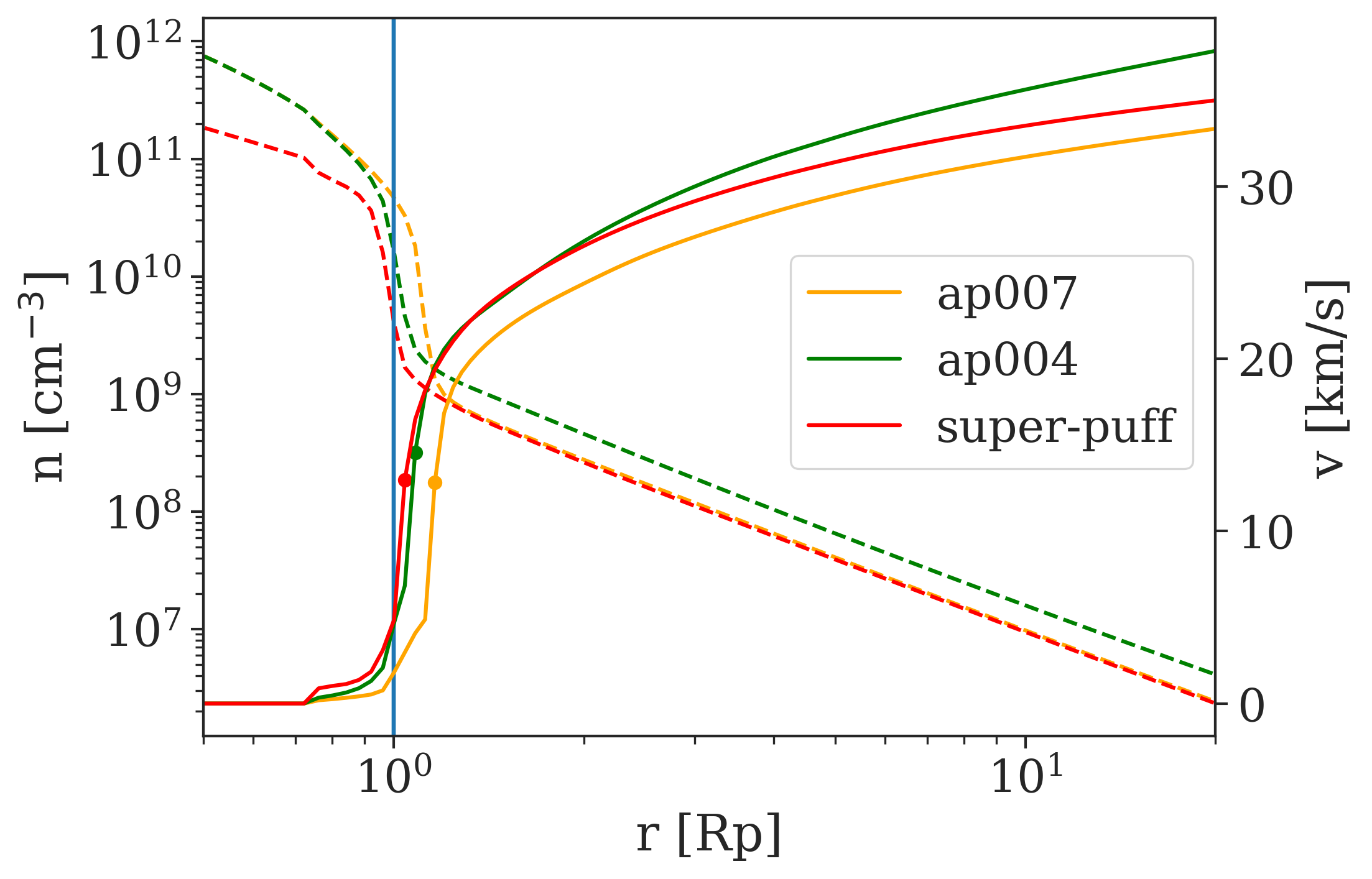}
\caption{Radial velocity and number density profiles of the outflow for all the models along the line connecting the planet and the star, centered at the planet. The triangle in the velocity lines marks the position of the sonic point. The solid vertical line indicates the R$_{\text{p}}=1$ position. Profiles are taken after $t=2.95$ days of evolution.}
\label{fig:radial_prof}%
\end{center}
\end{figure}   

In general, we found that photons are absorbed at a radius close to 1\,R$_\text{p}$, heating the planetary atmosphere and developing an outflow similar to a Parker-like wind. The outflow is characterized by a drop in density and an increase in velocity, becoming supersonic between 1 and 2\,R$_\text{p}$ in all models. This is visible in figure \ref{fig:radial_prof}, where the radial profiles of density and velocity for each model are shown. These profiles were taken from the outputs of the 3D simulations along the line connecting the star and the planet at t=1.5 days, after the simulation had reached a steady state.  

The different density and velocity profiles can be explained by differences in the F$_{\text{XUV}}$ that heat the planetary atmosphere and the planetary gravitational force. These two forces, thermal and gravitational, end up controlling the strength of the planetary outflow. 

\begin{figure*}[h!]
\begin{center}
\includegraphics[width=0.8\linewidth]{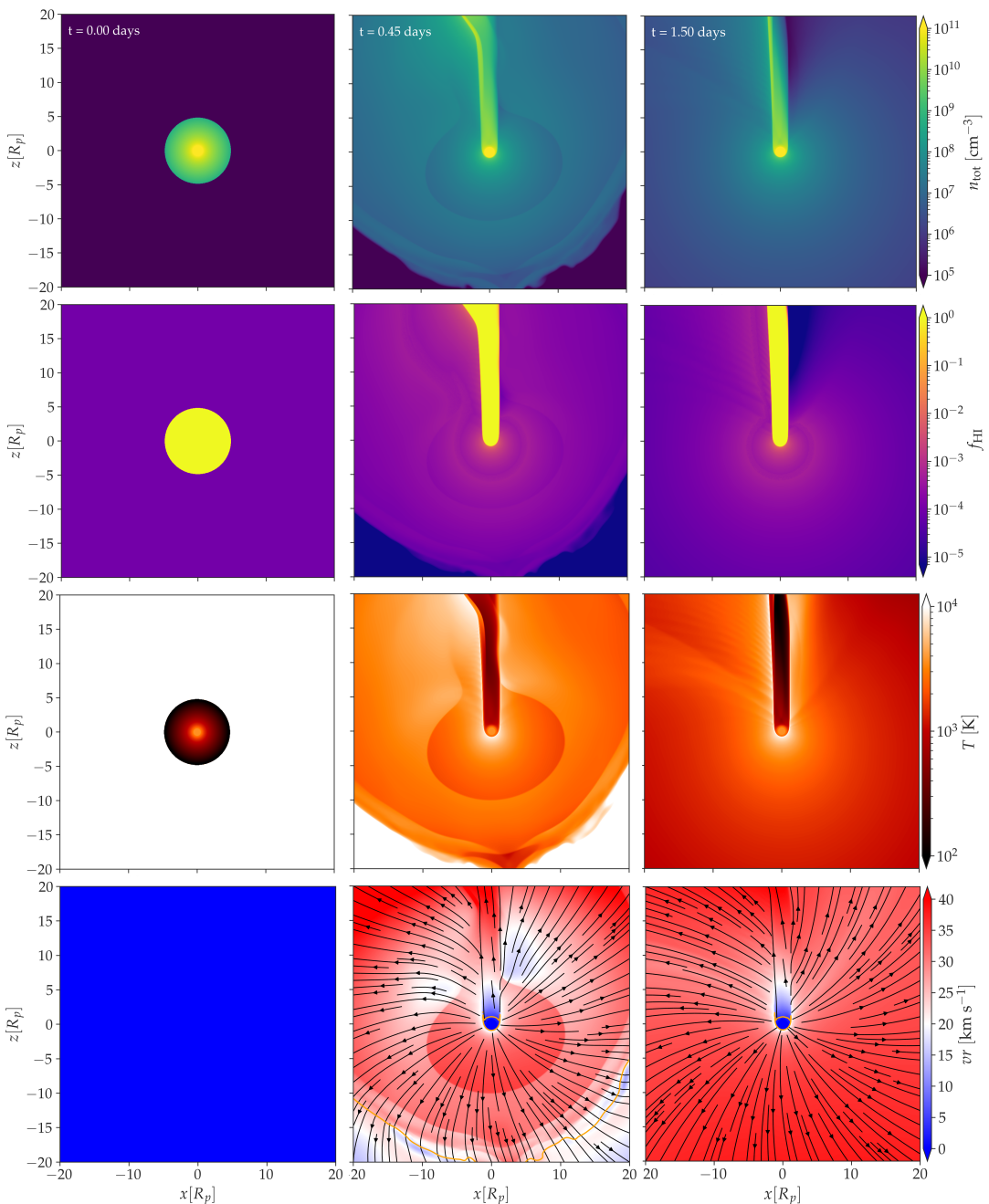}
\caption{Contour plots of total density (first row),  hydrogen neutral fraction (second row), temperature (third row), and radial velocity (module and direction, fourth row) in the orbital plane for model {\it super-puff} at different times during the simulation (different columns). 
The planet is at the center and the stellar radiation enters the grid from the $+z$ direction. The orange contour in the velocity plot shows where $v_r=c_s$. 
The planetary atmosphere evolves from hydrostatic a $t=0$ (first column) to a fully expanded planetary outflow at $t=1.5$ days (last column). Models {\it a$_{\text{p}}$004} and {\it a$_{\text{p}}$007} show a similar behavior in their evolution and this is why they are not plotted here. The planetary mass-loss rate found in all models is $\sim10^{12}$ g s$^{-1}$.}
\label{fig:contour}
\end{center}
\end{figure*}

Since models {\it a$_{\text{p}}$007} and {\it super-puff} have the planet at the same distance from the star, they receive the same amount of XUV flux and the difference between the two outflows lies in the gravitational force of the planet. Then, a higher velocity is achieved in the outflow from the super-puff planet. Model {\it a$_{\text{p}}$004} receives a higher XUV flux than model {\it a$_{\text{p}}$007} and this is why the velocity reaches a higher value despite sharing the same planetary characteristic. However, the differences appearing among these three outflows are not significant and the mass-loss rate found for each model is similar. 

The distribution of the outflow around the planet is shown in figure \ref{fig:contour} where the total density, fraction of neutral hydrogen, temperature and radial velocity are plotted from top to bottom in the orbital plane for the model {\it super-puff}. The sonic surface, marked with the orange contour in the radial velocity plot, is located near the planet's surface. The evolution of this model is shown across the columns of the same figure, ranging from $t=0$, when the planetary atmosphere is in static equilibrium, to $t\sim1.5$ days when the planetary outflow has expanded to the mesh boundaries and reached a quasi-steady state. Models {\it a$_{\text{p}}$004} and {\it a$_{\text{p}}$007} present a similar evolutionary behavior, and we decided not to include them. 

The escaping planetary material is asymmetrically distributed as a result of the incident stellar radiation and the gravitational pull of the star. On the night-side of the planet, where the stellar photons do not reach the atmosphere, the outflow is predominantly neutral with a small radial velocity ($<10$ km s$^{-1}$) forming a collimated, comet-like tail. On the day-side, the material escapes with a higher velocity, reaching $\sim40$ km s$^{-1}$ and is almost fully ionized due to the incoming stellar photons, as shown in the second row of figure \ref{fig:contour}. The escaped planetary material will reach the star in a path determined by stellar gravity and radiation, as there is no stellar wind to interact, likely forming a disk. This is further explored in the next section (\ref{Res1D}).

\begin{figure}[h!]
\begin{center}
\includegraphics[width=\linewidth]{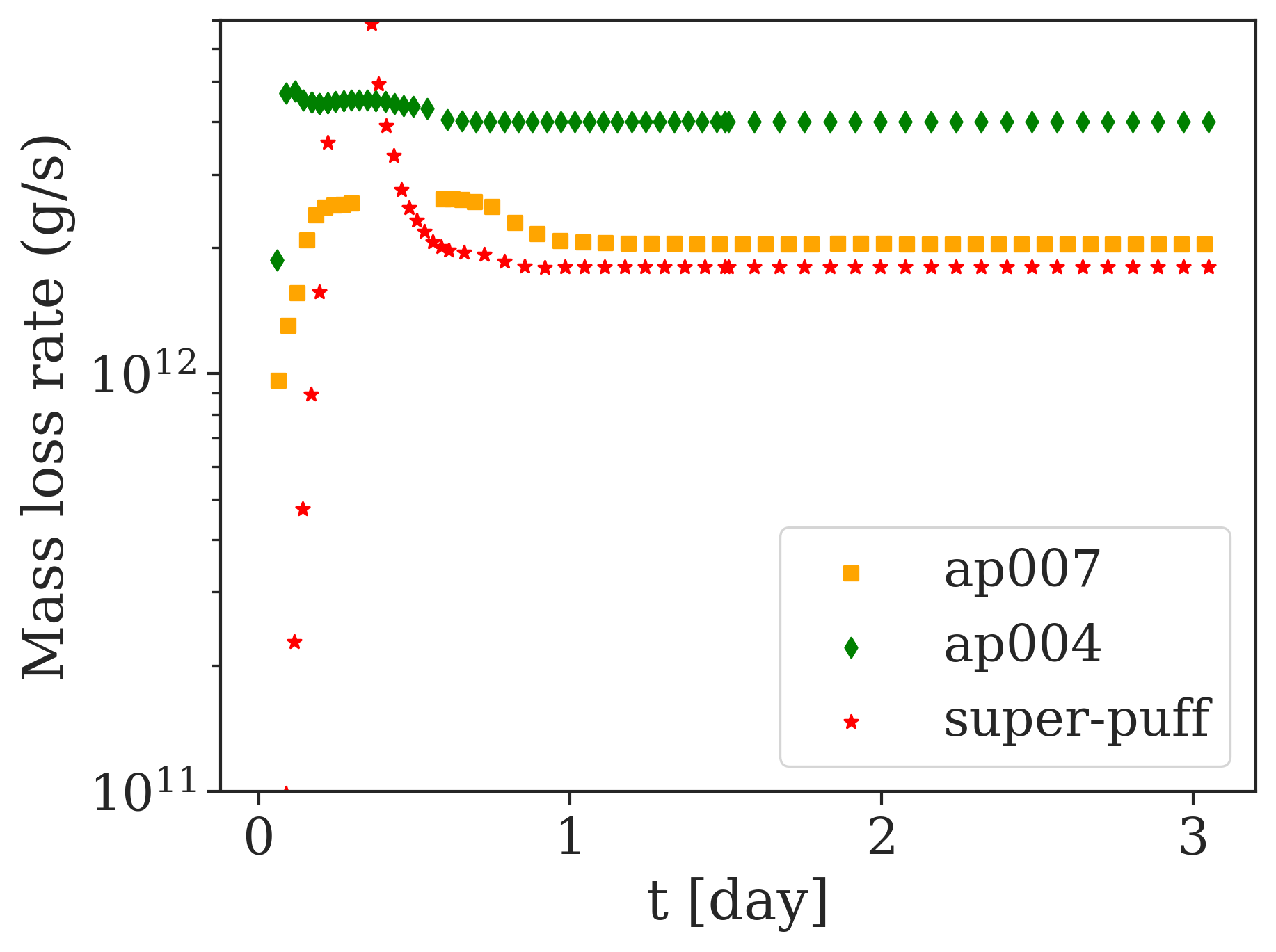}
\caption{Instantaneous planetary mass-loss rate computed every 500 steps of the simulation as a function of time for each model.}
\label{fig:mdot_time}%
\end{center}
\end{figure}

The mass-loss rate was calculated every 500 simulation steps for each model, as illustrated in figure \ref{fig:mdot_time}, as a function of time. The plot clearly shows that the instantaneous planetary mass loss rates quickly converge to an asymptotic value and remain stable.
The adopted value of $\dot{M}_{\text{p}}$ is the average over the last 20 instantaneous values and is presented in Table \ref{tab:2} for each model. For models {\it a$_{\text{p}}$004} and {\it a$_{\text{p}}$007}, the calculated mass loss rates are $4.01\times 10^{12}$ and $2.04\times10^{12}$ g s$^{-1}$, respectively. The mass-loss rate of the {\it super-puff} model is found to be $1.8\times10^{12}$ g s$^{-1}$, this smaller value resembles the slightly larger surface gravity of the planet with respect to the Neptune-size planet used in the other models.  
Considering that the highest uncertainty in our models are the values of the F$_\text{XUV}$, with an associated error of $\sim$30\%, we can conclude that in all cases, the mass loss rates are in agreement with the value estimated by \citet{Gaensicke2019} of $\sim5\times10^{11}$ g s$^{-1}$.

\subsection{1D disk evolution simulations: Mass accretion rates}\label{Res1D}
\begin{figure*}[ht!]
\begin{center}
\includegraphics[angle=0, width=0.9\linewidth]{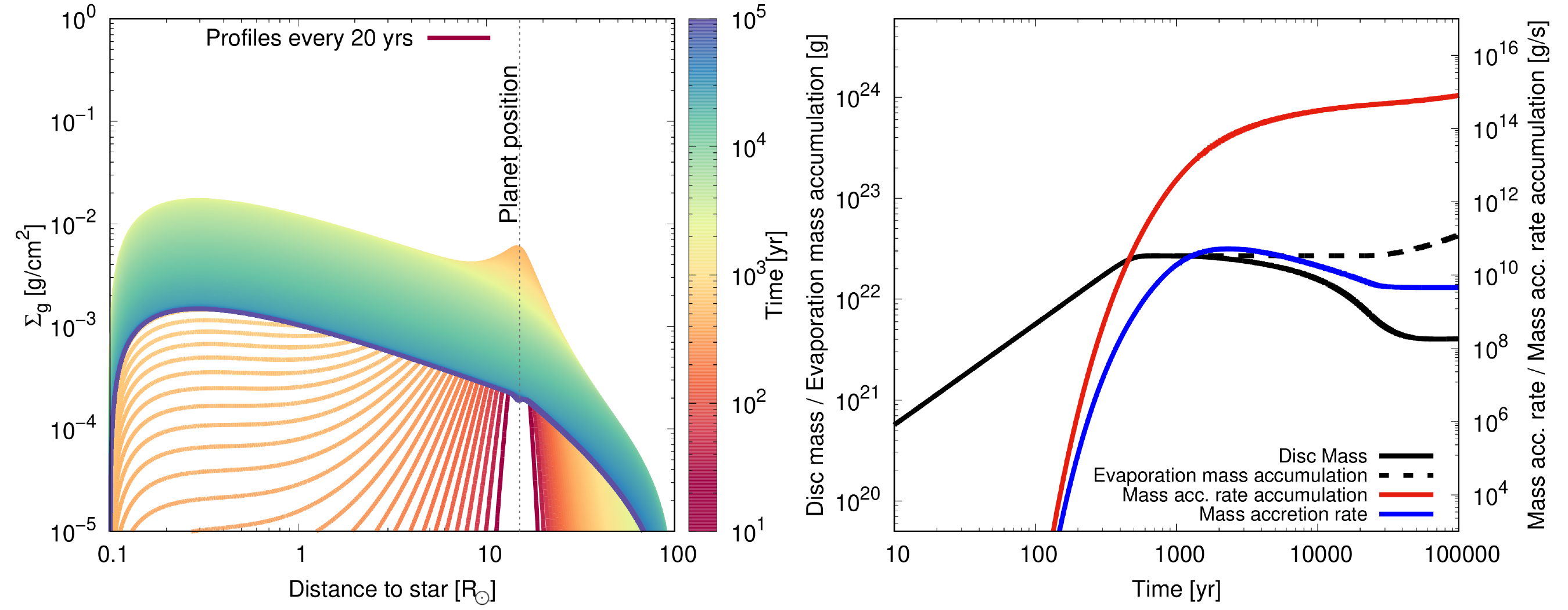}\\
\caption{Results for the 1D disk evolution simulation of case {\it super-puff}. The left panel shows the gas disk surface density profiles every  20 years for the {\it super-puff} planetary evaporated atmosphere, located at 15$R_{\odot} (0.07$~au) within a circular orbit. The right panel shows on the left y-axis the evolution of the mass of the disk (solid black line) and the evaporation mass accumulation attenuated by the disk shielding effect (dashed black line). The difference between these two lines is the accreted mass of the disk by the central star. Solid blue and red lines show the mass accretion rate and the mass accretion rate accumulated (right y-axis).}
\label{fig:1D}
\end{center}
\end{figure*}
The results of section \ref{Res3DLocal} suggest that the atmospheric material lost from the planet will reach the star. Hence, we investigated the time evolution of this material with our 1D viscous evolution model. For each of the three planetary cases, we performed a simulation considering a typical $\alpha-$viscosity disk parameter of $10^{-3}$ as in \citet{Metzger2012}. 

The results are shown in figure \ref{fig:1D} for the {\it super-puff} model, as the other two models do not show appreciable differences. On the left panel, the surface density profile plotted every 20 years shows the evolution of the disk. The injected planetary mass loss at the orbital distance of the planet initially forms a ring-like structure that evolves viscously, expanding both inward and outward to conserve angular momentum. To ensure that this gas remains gravitationally bound to the white dwarf, we verified at each timestep and radial bin that $(c_{\text{s}}/v_{\text{k}})^2 \ll 1$ \citep{Ida2020}. Although the planet exerts tidal torques onto the disk, these are insufficient to stop the gas from spreading beyond its orbital location. In Appendix \ref{App:tidal_torque}, we discuss and illustrate the fact that stopping the viscous spreading beyond the planetary orbit would require a planet of several Jupiter masses.

Within $\sim$200 years, the disk reaches the inner edge of our grid and gas accretion starts to become important. After $\sim$400 years, and despite the relatively low fraction of neutral hydrogen in the disk, assumed to be on the order of $10^{-4}$, as shown in the second row of fig. \ref{fig:contour}, the shielding effect of the gas disk onto the planet becomes effective. As a result, the planetary mass-loss rate decreases, leading to a decline in the disk mass until a balance between viscous evolution and (shielding-modulated) mass injection is established in less than $10^5$ years. At this stage, the disk mass in the {\it super-puff} case is of $3.39\times10^{21}$ g, while the accretion rate is $\sim4.25\times10^9$ g s$^{-1}$. Notably, this mass accretion rate is in agreement with the one observationally inferred by \citet{Gaensicke2019}, of $3.3\times10^9$ g s$^{-1}$. Although they share the same behavior, higher accretion rates were measured for the \textit{ap004} and \textit{ap007} cases, as shown in Table \ref{tab:2}. 

The right panel of figure \ref{fig:1D} shows the evolution of different disk parameters. It is easier to see from this figure that the disk mass diminishes when the gas in the disk starts to attenuate the XUV photons from the star around 200 years. It is also seen here that both the disk mass and mass accretion rate reach a constant value after 2$\times10^4$ years.
If mass injection were to completely stop because the planet has already lost its entire atmosphere, the disk mass would drop by more than two orders of magnitude in less than $\sim 1\times 10^5$ years. However, if the planet orbiting WD\,J0914 is indeed a Neptune-like planet with a typical envelope of 10\% of its total mass, the disk could remain in this steady state for several hundred thousand years.   

Due to the magnetic nature of white dwarfs and since we can use spectral observations of WD\,J0914 to place an upper limit of 100 kG on its magnetic field, we can further explore the effect that this magnetic field might have on the disc inner boundary. Specifically, we modeled a case were the disk is truncated at 1 R$_{\odot}$ and accretion proceeds along magnetic field lines. Even under these assumptions, we found no significant impact on the overall disk structure, evolution, or the inferred accretion rates.

\section{Discussion and conclusions}\label{sect_disk}

In this work, we modeled the atmospheric escape of WD\,J0914b using 3D radiative-hydrodynamic simulations. Although  analytical approximations such as those used by \citet{Gaensicke2019} provide a useful first-order estimate of atmospheric mass loss, they rely on some very specific assumptions; for example, the assumed heating efficiency or the radii where XUV irradiation is absorbed, which result in uncertainties of several orders of magnitude \citep{Kubyshkina2018, Krenn2021}. A more robust estimation of the planetary mass-loss rate can be obtained using hydrodynamic simulations. In our case, 3D models allow us to study the planetary mass loss as the natural response of the planetary atmosphere to the incoming stellar XUV flux and to investigate the distribution of the lost atmospheric material.

We also employed a 1D viscous model to simulate the long-term evolution of the gaseous disk formed by the evaporated planetary material, taking into account the attenuation that stellar XUV photons will suffer when traveling the disk. This allowed us to track the transport of gas towards WD\,J0914 and obtain self-consistent estimates of the mass accretion rates.

We considered different planet characteristics for this planetary system that could plausibly result from different dynamical scenarios \citep{VerasFuller2020}. Our simulations reveal that both Neptune-like and super-puff planets, at orbital distances between $0.04$ and $0.07$ au, undergo significant atmospheric escape driven by the intense stellar XUV radiation. In all modeled scenarios, the planet mass loss rates consistently reach $\sim10^{12}$ g s$^{-1}$, validating the analytical estimates of \citet{Gaensicke2019} and suggesting that the observational properties of WD\,J0914 are consistent with a range of planetary architectures.

Through our 3D simulations, we also examined the structure of the material escaping the planet. We found that on the day-side, the material expands at a velocity close to 40 km s$^{-1}$, while on the night-side, where stellar photons do not heat the atmosphere, the velocity is less than 10 km s$^{-1}$. This behavior is consistent across all models, with variations in the outflow characteristics depending on the amount of F$_{\text{XUV}}$ the planet receives and the strength of its gravity. Since there is no stellar wind to slow the expansion, the planetary outflow remains in the orbital path, ultimately reaching the star.

From our disk evolution models our results suggest that a Neptune-like or {\it super-puff} planet is likely embedded within the gaseous disk, which expands beyond the planetary orbit, as the tidal torques exerted by the considered planets are insufficient to counteract viscous spreading. In this scenario, the disk shielding effect becomes a critical mechanism by attenuating the irradiation reaching the planet, it regulates the mass-loss rate and limits both the total mass and the spatial extent of the disk.
For model {\it super-puff}, the disk reaches an equilibrium state in less than $10^5$ years. The computed accretion rate at this stage is $4.25\times10^{9}$ g~s$^{-1}$, in  good agreement with that derived in \citet{Gaensicke2019} from the spectroscopic absorption lines. Models {\it a$_p$004} and {\it a$_p$007} show  slightly higher results. 

 While our numerical results suggest a more extended disk than that reported by \citet{Gaensicke2019}, this apparent discrepancy likely reflects differences in the physical nature of the observational tracers. Outer disk radii inferred from forbidden emission lines are sensitive to local gas conditions and trace only those regions where density and temperature permit efficient excitation. These measurements therefore mark the extent of the line-emitting region rather than the full physical disk. At larger radii, declining density and temperature likely suppress collisionally excited emission, rendering the outer disk undetectable with current spectroscopic methods. This perspective is reinforced by recent analysis (Lagos-Vilches et al. in prep.), which suggests that a more detailed treatment of the line profiles supports a disk extending significantly beyond the initial estimates.

Even when our models only include hydrogen in the planetary atmosphere, any other element present in this atmosphere will follow (via collisions) the dynamics of hydrogen and escape the atmosphere in the same manner but in a smaller fraction. These heavier elements will be those found in the spectroscopic observations of \cite{Gaensicke2019}. The construction of a new set of models that simulates the 3D creation of the disk and allow for a more direct comparison with these spectroscopic observations is the next step in the study of this system.

\begin{acknowledgements}
We thank the referee for all comments and suggestions that improved our work and helped to better present the results.  
CVD and MPR gratefully acknowledge the support received through the Programa de Estadías Nacionales (PEN) from the Asociación Argentina de Astronomía, which fostered the development of this work. During the development of this work MPR was partially supported by PICT-2021-I-INVI-00161 from ANPCyT, Argentina and partially supported by PIP-2971 from CONICET (Argentina) and by PICT 2020-03316 from Agencia I+D+i (Argentina). 
MRS thanks for support from the research unit FOR2990 (eRO-STEP: Stellar endpoints with eROSITA) funded by the Deutsche Forschungsgemeinschaft and the internal research project PI\_LIR\_26\_07 at USM. O.T. acknowledges Proyectos Internos USM 2025-PI\_LII\_2025\_03 and FONDECYT iniciación 11260733.
MRS and MRP also acknowledge support from ESO/comit{\'e} mixto. CVD and MPR thank Octavio M. Guilera for valuable discussions that contributed to the improvement of this paper. The authors are grateful to Detlev Koester for sharing his model atmosphere code. This work used computational resources from CCAD-UNC, which is part of SNCAD-MinCyT, Argentina. This project has received funding from the European Research Council (ERC) under the European Union’s Horizon 2020 research and innovation programme (Grant agreement No. 101020057).

\textit{Software.} For this publication the following software packages have been used: 
\href{https://matplotlib.org/}{Python-matplotlib} by \citet{Hunter:2007}, \href{https://seaborn.pydata.org/}{Python-seaborn} by \citet{Waskom2021},
\href{https://numpy.org/}{Python-numpy} by \citet{harris2020array} and 
\href{https://pandas.pydata.org/}{Python-pandas}by \citet{reback2020pandas}.

\end{acknowledgements}

\bibliographystyle{aa}
\bibliography{Literature}

\begin{appendix}

\section{Effect of the planetary torques}
\label{App:tidal_torque}

In Sections \ref{sec:1Dmodel} and \ref{Res1D}, we describe our 1D model and 1D simulation results, respectively, particularly for the {\it super-puff} case. We argue that the torque exerted by the planet on the gas disk formed around the white dwarf is negligible due to the very low mass ratio. As a consequence, the disk is able to expand beyond the planet location and our results differ from those suggested by \citet{Gaensicke2019}.

Here, we illustrate this effect in a simplified manner. We considered the time evolution of a pre-existing disk extending from 0.1 to 10 R$_\odot$, of mass $M_\text{d}=4\times10^{22}$ g, which is approximately the maximum mass reached by the disk of our {\it super-puff} simulation described in section \ref{Res1D}. The disk is represented by an initial gas surface density profile given by 
\begin{equation}
    \Sigma_{\rm g} = \Sigma_{\rm{gas}}^{0} \left(\frac{R}{{R}_{\rm c}}\right)^{-\gamma} e^{-(R/{R}_{\rm c})^{2-\gamma}}
    \label{equ:densi_inicial},
\end{equation}
with $\Sigma_{\rm{gas}}^{0}= (2-\gamma) \frac{{M}_{\rm d}}{2\pi {{R}_{\rm c}}^2}$ a normalization constant, $\gamma=1$ and R$_{\rm c}=10$ R$_\odot$ the disk characteristic radius. To isolate the role of the planetary torque, we neglect both the atmospheric escape and gas injection from the planet, focusing solely on the interaction between the planet and the disk. To maximize the potential impact of the torque, we adopt $f=1$ in eq. \ref{eq:evol_gas}.

\begin{figure}[ht!]
\begin{center}
\includegraphics[angle=0, width=0.85\linewidth]{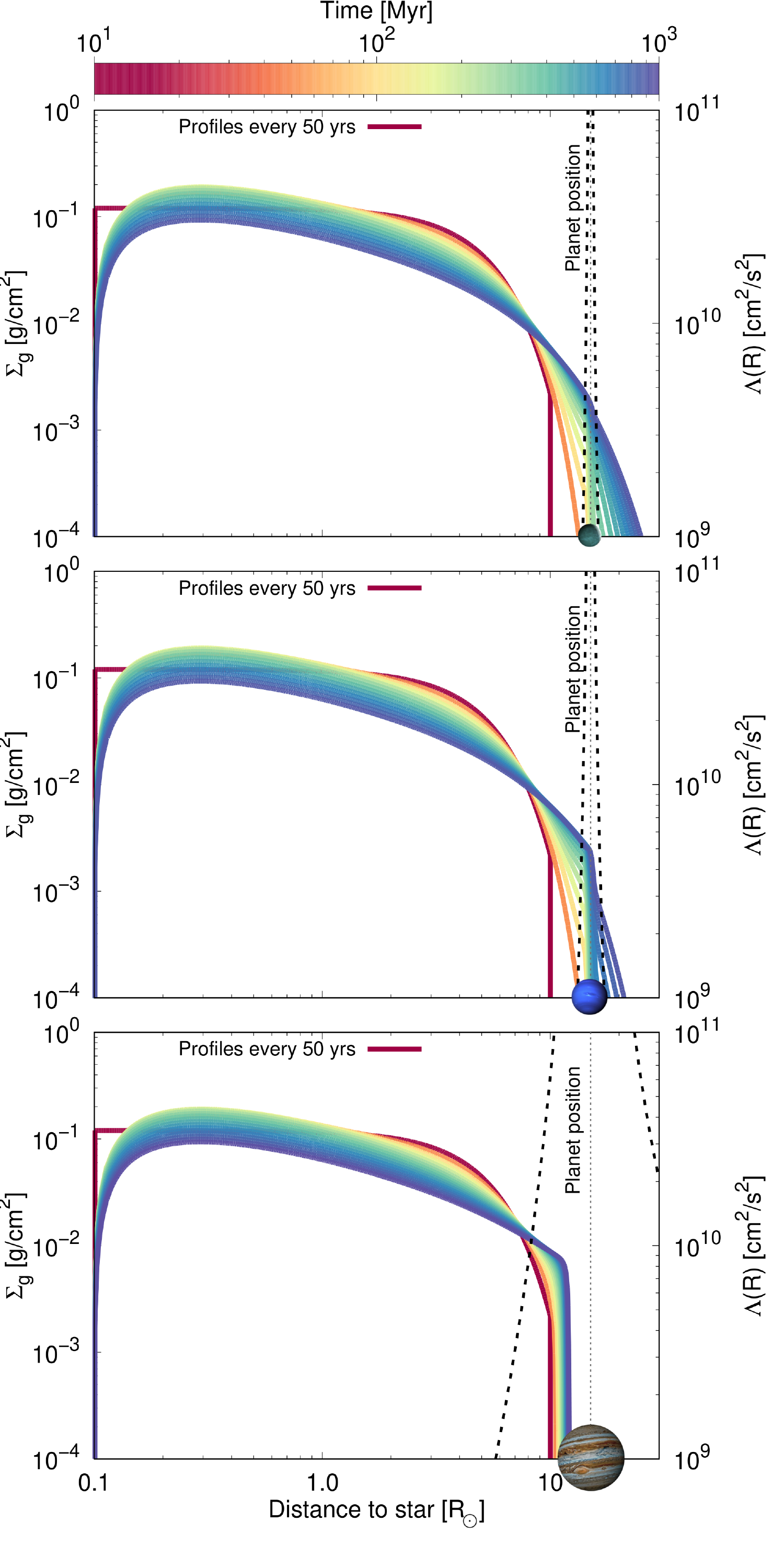}\\
\caption{Time evolution of a gas disk extended from 0.1 to 10 R$_\odot$ orbiting a white dwarf together with a super-puff planet of 6 M$_\odot$ (left), a Neptune-like planet of 18 M$_\odot$ (middle) and a Jupiter-like planet of 5 M$_\text{J}$ (right). The initial mass of the disk, M$_{\text{d}}=4\times10^{22}$ g, is the same for the three cases. The magnitude of the specific tidal torque $\Lambda(R)$ is represented in the right y-axis.}
\label{fig:App1D}
\end{center}
\end{figure}

Figure~\ref{fig:App1D} shows the evolution of the disk under the influence of planets with different masses. In the left panel, the planet is a {\it super-puff} with 6 M$_\oplus$ (our base case); in the middle panel, a Neptune-like planet similar to HAT-P-26 (18 M$_\oplus$); and in the right panel, a five-Jupiter-mass planet. Although the simulations span only 1000 years, the dependence of the torque effect on the mass ratio $q$ is clearly evident.

For the {\it super-puff} and Neptune-mass cases, planetary torques are insufficient to stop viscous spreading, and the disk expands beyond the planetary orbit, indicating that such planets are likely embedded within the gas disk. In contrast, the five-Jupiter-mass planet significantly alters the disk evolution and is able to inhibit its outward expansion. The black dashed lines indicate the magnitude of $\Lambda(R)$ exerted by each planet, for comparison.

\end{appendix}

\end{document}